\documentclass[a4paper,12pt]{article}
\usepackage[utf8]{in
    putenc}
\usepackage{cancel}
\usepackage{ulem}
\usepackage{amsfonts}
\usepackage{amssymb}
\usepackage{graphicx}
\usepackage{amsmath,bm}
\usepackage{enumerate}
\usepackage{mathtools}
\usepackage{tikz} \usetikzlibrary{calc}
\usepackage[countmax]{subfloat}
\usetikzlibrary{fit}
\usepackage{xcolor}
\usepackage{float}
\usepackage[symbol]{footmisc}
\usepackage{color}
\usepackage{here}
\usepackage{cite}
\usepackage{subcaption}
\usepackage{mwe}
\usepackage[outline]{contour}
\usepackage{tabularx}
\usepackage{multirow}
\usepackage{mathrsfs}
\usepackage{float,epsfig}
\usepackage{dcolumn}
\usepackage{pgfplots}
\pgfplotsset{compat=1.18}
\usepackage{graphicx}
\usepackage{bm}
\usepackage{amsmath,amssymb,amsthm}
\usepackage[colorlinks=true,linkcolor=blue,citecolor=red]{hyperref}
\definecolor{lime}{HTML}{A6CE39}
\newcommand{\orcidicon}{%
    \begin{tikzpicture}
    \draw[lime, fill=lime] (0,0)
        circle [radius=0.16]
        node[white] {{\fontfamily{qag}\selectfont \tiny ID}};
    \draw[white, fill=white] (-0.0625,0.095)
        circle [radius=0.007];
    \end{tikzpicture}   \hspace{-2mm}
}

\newcommand\orcidHasan{{\href{https://orcid.org/0000-0001-7408-0910}{\orcidicon}}}
\newcommand\orcidKarima{{\href{https://orcid.org/0000-0001-5419-8516}{\orcidicon}}}
\newcommand\orcidFaical{{\href{https://orcid.org/0000-0002-2977-0821}{\orcidicon}}}

\title{\bf Nonextensive Black Hole Thermodynamics from Generalized Euclidean Path Integral and Wick's Rotation}

\author{
F. Barzi\orcidFaical\!\!$^{1,3}$\thanks{faical.barzi@edu.uiz.ac.ma},  
 H.  El Moumni\orcidHasan\!\!$^1$\thanks{h.elmoumni@uiz.ac.ma (Corresponding author)}, K. Masmar\orcidKarima\!\!$^{1,2}$\thanks{karima.masmar@gmail.com}
\\
{\small $^{1}$ LPTHE, Physics Department, Faculty of Sciences, Ibnou Zohr University, Agadir, Morocco. }\\
{\small $^{2}$Laboratory of  High Energy Physics and Condensed Matter
HASSAN II University,}\\{\small Faculty of Sciences Ain Chock, Casablanca, Morocco.}\\
{\small $^{3}$CRMEF, Regional Center for Education and Training Professions Marrakesh, Morocco.}
}
\date{\today}
\begin{document} 
\maketitle
\vspace{-1cm}
\begin{abstract}
This paper extends the Euclidean path integral formalism to account for nonextensive statistical mechanics. Concretely, we introduce a generalized Wick's rotation from real time $t$ to imaginary time $\tau$ such that, $t\rightarrow-i f_\alpha(\tau)$, where $f_\alpha$ a differentiable function and $\alpha$ is a parameter related to nonextensivity. The standard extensive formalism is recovered in the limit $\alpha\rightarrow0$ and  $f_0(\tau)=\tau$. Furthermore, we apply this generalized Euclidean path integral to black hole thermodynamics and derive the generalized Wick's rotations given the nonextensive statistics. The proposed formulation enables the treatment of nonextensive statistics on the same footing as extensive Gibbs-Boltzmann statistics. Moreover, we define a universal measure, $\eta$, for the nonextensivity character of statistics. Lastly, based on the present formalism, we strengthen the equivalence between the AdS-Schwarzschild black hole in Gibbs-Boltzmann statistics and the flat-Schwarzschild black hole within R\'enyi statistics and suggest a potential reformulation of the $AdS_5$/$CFT_4$ duality. 
{\noindent}
\end{abstract}

\tableofcontents
\section{Introduction}
\paragraph{} The path integral formulation of quantum mechanics, developed in the mid-twentieth century \cite{Feynman:1948ur,feynman2010quantum}, stands as a remarkable synthesis of key theoretical physics concepts and serves as a powerful computational tool. This technique has been applied to a wide range of physical systems across diverse contexts, including quantum mechanics\cite{FEYNMAN2005}, quantum field theory\cite{maggiore2005modern,Fai:2019llg}, gauge field theory\cite{faddeev1993gauge,chaichian2012introduction}, black hole physics\cite{PhysRevD.13.2188}, quantum gravity \cite{hawking78}, string theory \cite{Polchinski:1998rq,ISHIBASHI2000149}, topology \cite{MouroAspectsOT,doi:10.1142/9789814415330_0004}, condensed matter physics \cite{PhysRevB.30.2555,PhysRevLett.73.2145,PhysRevLett.124.116401}, and optical communications \cite{Reznichenko_2017}, among others \cite{10.1063/1.466438,PhysRevE.47.118,PhysRevLett.72.1341}. Indeed, in statistical physics, path integrals laid the foundation for the first formulation of the renormalization group transformation and are widely used to study systems with random impurity distributions \cite{Imai_2007}. In particle physics, they have been crucial in understanding and accounting for the presence of instantons \cite{PhysRevD.53.6979}. Quantum field theory benefits from path integrals as the natural framework for quantizing gauge fields. In chemical, atomic, and nuclear physics, this approach has been applied to various semiclassical schemes in scattering theory. Moreover, path integrals offer a powerful means to explore classical and quantum fields' topological and geometrical properties, facilitating novel perturbative and non-perturbative analyses of fundamental natural processes \cite{Green_Schwarz_Witten_2012}.

\paragraph{}The Boltzmann-Gibbs (BG) statistics has long been the cornerstone for describing a broad class of physical systems, providing over a century of successful applications\footnote{\tt This year's Nobel Prize in Physics honors groundbreaking research on the application of Boltzmann statistics to neural networks and machine learning.}. It is particularly effective for systems with predominantly chaotic dynamics, such as classical systems exhibiting mixing, ergodicity, and a positive maximal Lyapunov exponent. However, many complex physical systems fall outside the scope of this framework, especially those where the maximal Lyapunov exponent vanishes, indicating a departure from simple chaotic behavior. To better describe the statistical properties of such systems, various generalized forms of statistical mechanics have emerged, including nonadditive entropies, Kappa-distributions \cite{Pierrard2010,Livadiotis2013}, $q$-Gaussians \cite{diaz2005qqw,diaz2009wqq}, and Superstatistics \cite{Beck2003,Beck2017,Beck2020,Hanel2011eee}. These generalized approaches have demonstrated a wide range of applications both within and beyond physics.

\paragraph{}For the building and discussion of relativistic quantum field theoretical models, the concept of \textit{passage to imaginary time} has proven to be an invaluable technique. This approach was initially introduced by Dyson\cite{Dyson49} and later formalized by Wick\cite{wick1954}, giving rise to the well-known \textit{Wick rotation}. Building on this foundation, Schlingemann\cite{schling99} provided a more rigorous framework that connected Euclidean and Lorentzian quantum field theories, leveraging the Osterwalder-Schrader theorem\cite{Oster1973, Oster75}. This theorem outlines the necessary and sufficient conditions for a consistent transition between these frameworks.

Applying a Wick rotation to the Lorentzian path integral reveals its close resemblance to the partition function in statistical mechanics. Specifically, the Euclidean path integral sums over all possible paths, with each path weighted by an effective energy-like term derived from the action in imaginary time—mirroring how the partition function sums over all states of a system, weighted by their respective energies. This formalism in Quantum Mechanics (\textit{QM}) stems from the time-evolution operator $\mathcal{U}(t)$, which satisfies the following differential equation
\begin{equation}\label{eq_1}
i\hbar\frac{\partial \mathcal{U}}{\partial t}=H \mathcal{U}.
\end{equation}
Where $H$ is the Hamiltonian operator. The solution of the above equation for a time-independent Hamiltonian reads as
\begin{equation}\label{eq_2}
\mathcal{U}(t)=\exp\left(\frac{-i H t}{\hbar}\right).
\end{equation}
Probability amplitudes are given by matrix elements of $\mathcal{U}(t)$. Wick's rotation permits a connection with the partition function of statistical mechanics such that by analytical continuation to imaginary time $t\rightarrow -i\tau$ we get,
\begin{equation}\label{eq_3}
\mathcal{U}(\tau)=\exp\left(\frac{- H \tau}{\hbar}\right).
\end{equation}
Which is a solution to the diffusion-like differential equation,
\begin{equation}
\frac{\partial \mathcal{U}}{\partial \tau}=-\frac{H \mathcal{U}}{\hbar}
\end{equation}
Then, the partition function $Z[\beta]$ is given by,
\begin{equation}
Z[\beta]=tr\left[\exp\left(-\beta H\right)\right]=\oint Dg\,\exp\left(-\mathcal{I}_E[g]\right),
\end{equation}
where $tr[\,.\,]$ is the trace operator,  $\displaystyle\beta=\frac{\tau}{\hbar}$, and $\mathcal{I}_E$ is the Euclidean action. The integration is performed over all closed trajectories in phase space. We readily get the thermodynamic free energy from the partition function as,
\begin{equation} 
F=U-TS=-\frac{1}{\beta}\ln\left(Z[\beta]\right).
\end{equation}

Likewise, all other relevant thermodynamic quantities can be computed from $Z[\beta]$.


\paragraph{}The setup demonstrates that the exponential form of the time-evolution operator in Eq.\eqref{eq_2} aligns with the Boltzmann exponential probability factor in Eq.\eqref{eq_3} after applying a Wick rotation. This connection suggests that the extensive nature of Gibbs-Boltzmann statistics is fundamentally tied to the linearity of the Schrödinger equation—without which, the exponential form of the time-evolution operator would not be feasible. In contrast, nonextensive statistical mechanics modifies the partition function away from its exponential form, typically by introducing new functional forms and parameters to capture nonextensive behavior. However, the linearity of the Schrödinger equation remains unaltered in such approaches, creating a fundamental inconsistency.  This inconsistency arises because, while the statistical mechanics framework attempts to generalize Gibbs-Boltzmann statistics, the quantum mechanical side lacks any modification to justify this shift. This discrepancy presents a theoretical gap that is unsatisfactory for a coherent understanding of the relationship between quantum mechanics and nonextensive statistical mechanics. Thus, there is a clear need for a consistent theoretical foundation that links the path integral formulation of quantum mechanics with nonextensive statistical approaches. In this paper, we aim to address this imbalance and propose a more cohesive framework.

 \paragraph{}This paper is structured as follows: In Sec.\ref{sec2}, we introduce the generalized Euclidean path integral formalism alongside the extended version of Wick's rotation. Sec.\ref{sec3} demonstrates the application of this new formalism to derive nonextensive black hole thermodynamics, achieving a comparable status to Boltzmannian statistics. Additionally, we introduce a measure of nonextensivity to quantify the statistical character. Sec.\ref{sec4} explores the proposed \textit{R\'enyi/AdS} equivalence and suggests a potential reformulation of the $AdS_5/CFT_4$ duality. A general discussion and concluding remarks are provided in Sec.\ref{sec5}.

 \section{Generalized Euclidean path integral}\label{sec2}
We address this discrepancy by extending the existing procedure that connects the two frameworks: \textit{Wick's rotation}. Specifically, we introduce a \textit{generalized Wick's rotation}, offering a more comprehensive approach to bridge the gap between the path integral formulation of quantum mechanics and nonextensive statistical mechanics as
\begin{equation}\label{eq_7}
t\longrightarrow-i f_\alpha(\tau).
\end{equation}

Where $\alpha$ is a real parameter related to nonextensivity such that, as $\alpha\rightarrow0$, we recover the extensive Wick's rotation $t\longrightarrow-i\tau$, that is $\underset{\alpha\rightarrow0}{\lim}\,f_\alpha(\tau)=\tau$. Applying the generalized rotation, Eq.\eqref{eq_7}, to Eq.\eqref{eq_1}, we get
\begin{equation} 
\frac{\partial \mathcal{U_\alpha}}{\partial \tau}=-\frac{H \mathcal{U_\alpha}}{\hbar f^{'}_\alpha(\tau)}.
\end{equation}

Where $ f^{'}_\alpha(\tau) $ is the first derivative of $f_\alpha(\tau)$. The generalized rotated operator $\mathcal{U_\alpha}$ is then given by
\begin{equation}\label{eq_9}
\mathcal{U_\alpha}(\tau)=\exp\left(-\frac{H}{\hbar}\int_0^\tau\frac{d\tilde{\tau}}{f^{'}_\alpha(\tilde{\tau})}\right)=\exp\left(-H\int_0^\beta\frac{d\tilde{\beta}}{k_\alpha(\tilde{\beta})}\right).
\end{equation}
Here we put $k_\alpha(\beta)=f^{'}_\alpha(\tau)$. The partition function is therefore generalized as
\begin{equation} 
Z_\alpha[\beta]=tr\left[\mathcal{U_\alpha}\right]=\oint Dg\,\exp\left(-\mathcal{I}^\alpha_E[g]\right).
\end{equation}
Herein, $ \mathcal{I}_E^\alpha $ is the nonextensive Euclidean action derived from the Lorentzian action $\mathcal{I}_L$ through the transformation Eq.\eqref{eq_7} such that
\begin{equation} 
i\mathcal{I}_L=i\int dt\, L(t)=-\int d\tau f^{'}_\alpha(\tau)\, (-L(\tau))=-\int d\tau \, (-L^\alpha(\tau))=-\mathcal{I}_E^\alpha.
\end{equation}
 Where $ L^\alpha(\tau)= f^{'}_\alpha(\tau)\, L(\tau)$ is the generalized Euclidean Lagrangian. With the above-extended formalism, we reconcile the linearity of \textit{QM} through the Schrodinger equation with the inherent non-linearity of the nonextensive statistical mechanics as proposed by Tsallis and al.\cite{Tsallis1988,Tsallis1998,Tsallis2004,Umarov2022,Tsallis2023}.

 
 \paragraph{}The free energy for the nonextensive statistics is computed as
\begin{equation}\label{eq_12}
F_\alpha=U_\alpha- \frac{S_\alpha}{\beta} =-\frac{1}{\beta}\ln\left(Z_\alpha\right).
\end{equation}

Here, $S_\alpha$ represents the nonextensive entropy, and $\displaystyle \beta=\frac{1}{T}=\frac{\partial S_\alpha}{\partial U_\alpha}$ denotes the inverse temperature in the framework of nonextensive thermodynamics. The choice of the nonextensive entropy, $S_\alpha$, determines the functional form of $f_\alpha(\tau)$ through Eq.\eqref{eq_12}, as will be demonstrated below. Using the well-known saddle point semiclassical approximation, the partition function $Z_\alpha$ takes the following form
\begin{equation}\label{eq_13}
Z_\alpha\approx\exp\left(-\mathcal{I}_E^\alpha[g_{cl}]\right),
\end{equation}

where $ g_{cl} $ is the classical solution of the system's Euler-Lagrange equations. One then has
\begin{align} 
&F_\alpha=U_\alpha-\frac{S_\alpha}{\beta} \approx\frac{\mathcal{I}_E^\alpha[g_{cl}]}{\beta}.\\\label{eq_15}
&\implies \beta U_\alpha- S_\alpha =\mathcal{I}_E^\alpha[g_{cl}]\equiv\mathcal{I}^\alpha_{cl} .
\end{align}
Eq.\eqref{eq_15} is the condition to determine $f_\alpha(\tau)$ and the required Wick's rotation. Explicitly, we write for a given spacetime metric $g$, ($U_\alpha=M$),
\begin{equation}\label{eq_16}
M\frac{\partial S_\alpha}{\partial M}- S_\alpha=-\int_{0}^{\tau}d\tilde{\tau}\, L^\alpha[g_{cl}]
\end{equation}

Where $ L^\alpha[g_{cl}]\equiv L^\alpha_{cl}$ is the Euclidean Lagrangian computed at the saddle point $g_{cl}$.\\

It is worth noting that the choice of $f_\alpha(\tau)$ could affect the Euclidean-invariance of the rotated Lagrangian, $L^\alpha_{cl}$, and thus of its Euclidean action, which may supplement further restrictions on the set of possible functions $f_\alpha$, if such invariance is to be preserved. Nonetheless, the Lorentzian-invariance is the physical invariance and it is still conserved. Additionally, since Wick's rotation function is derived from the generalized entropies, it is evident that they should inherit the consequences of some or all of the Shannon-Khinchin axioms\cite{Shannon1948,Khinchin1957}. 
\paragraph{}It is seen from Eq.\eqref{eq_9} that the simplest choice which is Euclidean-invariant, corresponds to the obvious extensive GB case, $f^{'}_0(\tau)=k_0(\beta)=1$. 
A close inspection of Eq.\eqref{eq_16} gives a formal expression for $k_\alpha(\beta)\equiv f^{'}_\alpha(\tau)$ defined in Eq.\eqref{eq_9}, in the general case,
\begin{align}\label{eq_17}
&\int_{0}^{\beta}d\tilde{\beta}\, k_\alpha(\tilde{\beta})\,L_{cl}(\tilde{\beta})=S_\alpha-M\frac{\partial S_\alpha}{\partial M}\\
\implies &k_\alpha(\beta)\,L_{cl}(\beta)=\frac{\partial}{\partial \beta}\left(S_\alpha-M\frac{\partial S_\alpha}{\partial M}\right).
\end{align}
That is,
\begin{align}
k_\alpha(\beta)&=\frac{1}{L_{cl}(\beta)}\frac{\partial}{\partial \beta}\left(S_\alpha-M\frac{\partial S_\alpha}{\partial M}\right)\\
&=\frac{1}{L_{cl}(\beta)}\left(\frac{\partial S_\alpha}{\partial \beta}-\frac{\partial( M\beta)}{\partial \beta}\right)\\\label{eq_21}
&=\bm{\frac{1}{L_{cl}(\beta)}\left(\frac{\partial S_\alpha}{\partial \beta}-\beta\frac{\partial M}{\partial \beta}-M\right)}
\end{align}

\paragraph{} When a generalized Wick's rotation for a given nonextensive statistics is calculated using Eq.\eqref{eq_21}, it is placed on equal footing as the Gibbs-Boltzmann extensive statistics which also obeys in the present generalization,
\begin{equation} 
k_0(\beta)=1=\frac{1}{L^0_{cl}(\beta)}\left(\frac{\partial S_0}{\partial \beta}-\beta\frac{\partial M}{\partial \beta}-M\right),
\end{equation}
where $S_0\equiv S_{GB}$, is the Gibbs-Boltzmann entropy. Thus,
\begin{equation}\label{eq_23}
L^0_{cl}(\beta)=\frac{\partial S_{GB}}{\partial \beta}-\beta\frac{\partial M}{\partial \beta}-M.
\end{equation}
We employ the notation \textit{index} "$0$" to denote the extensive Boltzmannian statistics, which corresponds to the case where the nonextensive parameter vanishes $(\alpha = 0)$.

\section{Applications to black hole thermodynamics}\label{sec3}

\paragraph{}One of the main motivations for deriving nonextensive statistics from the generalized Euclidean path integral is the need to introduce nonextensive entropies into black hole thermodynamics in order to account for their nonextensive nature. Through the standard Euclidean path integral formulation, where Wick's rotation function is limited to $f_0(\tau)=\tau$, one can uniquely obtain the Hawking temperature as the thermodynamic temperature of the black hole.   A challenge arises when attempting to modify the black hole thermodynamic statistics away from the Gibbs-Boltzmann framework, as altering the entropy inevitably leads to a corresponding change in temperature. This contradicts the traditional understanding of Hawking temperature, resulting in a tension between the need for nonextensive behavior in black holes and the unique nature of Hawking temperature as derived from Wick’s rotation. The solution proposed in this study addresses this issue by introducing a generalized Wick's rotation, which produces a temperature that aligns with the chosen statistical framework for the black hole system. In this view, the Hawking temperature is no longer unique; it represents the temperature that aligns specifically with Boltzmannian statistics. Thus, it becomes \textit{equally} possible to apply any statistical model to a given black hole without encountering contradictions.

\paragraph{} In this section, we assess the applicability of our formalism by analyzing a range of well-known black hole thermodynamic systems, considering both extensive and nonextensive statistical frameworks. Furthermore, we propose a novel universal measure for quantifying the degree of nonextensivity in a given statistical model, derived from the generalized Wick's rotation introduced in the preceding section.

\subsection{The \textit{4d}-AdS Schwarzschild black hole in Gibbs-Boltzmann statistics}
Let's apply Eq.\eqref{eq_23} to the case of the four-dimensional asymptotically-AdS Schwarzschild black hole within GB-statistics ($4d$ AdS-Sch). We have for the entropy, mass, and horizon radius of this black hole \footnote[2]{The horizon radius is given in terms of the cosmological constant $\Lambda<0$ and the inverse temperature $\beta$, pending the condition that the AdS-Sch black hole phase can appear from the thermal radiation phase. This holds for $\displaystyle\beta^2<-\frac{4\pi^2}{\Lambda}=\beta_{min}^2$\cite{Wei2020}. Similar remarks are true also for subsequent black hole systems.},
\begin{equation} 
S_{GB}=\pi r_h^2, \quad M=\frac{r_h}{2}-\frac{\Lambda r_h^3}{6}\quad \text{and}\quad  r_h=\displaystyle \frac{ \sqrt{\Lambda \beta^{2} + 4 \pi^{2}} - 2 \pi}{\Lambda \beta},
\end{equation}
here $\Lambda<0$ is the cosmological constant associated with the AdS spacetime radius via $\Lambda=\frac{3}{\ell^2}$. By substitution in Eq.\eqref{eq_23}, the Euclidean gravitational Lagrangian $ L^0_{cl}(\beta,\Lambda) $ is found to be
\begin{equation} 
 L^0_{cl}(\beta,\Lambda)=\displaystyle\frac{4 \pi ^2 \left(\beta ^2 \Lambda -4 \pi  \sqrt{\beta ^2 \Lambda +4 \pi ^2}+8 \pi ^2\right)-\beta ^4 \Lambda ^2}{3 \beta ^3 \Lambda ^2 \sqrt{\beta ^2 \Lambda +4 \pi ^2}},
\end{equation}
or equivalently in terms of the imaginary time $\tau=\beta$ ($\hbar=1$)
\begin{equation}\label{eq_26}
L^0_{cl}(\tau,\Lambda)=\displaystyle \frac{4 \pi ^2 \left(\Lambda  \tau ^2-4 \pi  \sqrt{\Lambda  \tau ^2+4 \pi ^2}+8 \pi ^2\right)-\Lambda ^2 \tau ^4}{3 \Lambda ^2 \tau ^3 \sqrt{\Lambda  \tau ^2+4 \pi ^2}}.
\end{equation}

Integrating Eq.\eqref{eq_26} gives the on-shell gravitational Euclidean action,
\begin{equation} 
\mathcal{I}^0_{cl}(\Lambda)=\displaystyle\frac{\Lambda \tau ^2+2 \pi  \left(\sqrt{\Lambda \tau ^2+4 \pi ^2}+4 \pi \right)}{3 \Lambda \left(\sqrt{\Lambda \tau ^2+4 \pi ^2}+2 \pi \right)}-\frac{\pi }{\Lambda }.
\end{equation}
Here $ \mathcal{I}^0_{cl}(\Lambda) $ is finite since it is the sum of the bulk action which diverges because spacetime has infinite volume, the Gibbon-Hawking action for the boundary contribution and the counterterm action to cancel divergences. For small cosmological constant, $ \Lambda\ll1 $, we get
\begin{align} 
&L^0_{cl}(\tau,\Lambda)=\displaystyle - \frac{\tau}{8 \pi} + \frac{\Lambda \tau^{3}}{96 \pi^{3}} + O\left(\Lambda^{2}\right),\\
&\mathcal{I}^0_{cl}(\Lambda)=\displaystyle  \frac{\tau^{2}}{16 \pi} - \frac{\Lambda \tau^{4}}{384 \pi^{3}} + O\left(\Lambda^{2}\right).\label{eq_29b}
\end{align}

The associated gravitational partition function for small $\Lambda\ll1$ is given through Eq.\eqref{eq_13} such as 

\begin{equation} 
Z_0(\Lambda)=\exp\left(-\frac{\beta^{2}}{16 \pi} + \frac{\Lambda \beta^{4}}{384 \pi^{3}}\right).
\end{equation}

Since $\Lambda$ is negative, the partition function remains bounded. In the limit where the cosmological constant vanishes, $\Lambda \rightarrow 0$, we recover the well-known results for the Gibbs-Boltzmann statistics: the 4-dimensional asymptotically flat Schwarzschild classical Lagrangian, $L_{cl}$, the classical action, $\mathcal{I}^0_{cl}$, and the gravitational partition function, $Z_0$, as follows
\begin{equation}\label{eq_31b}
    L_{cl}(\tau)=\displaystyle-\frac{\tau}{8\pi}, \quad \mathcal{I}^0_{cl}(\tau)=\displaystyle\frac{\tau^2}{16\pi},\quad \text{and}\quad Z_{0}(\beta)=\displaystyle\exp\left({-\frac{\beta^2}{16\pi}}\right).
\end{equation}

We note that Eqs.\eqref{eq_31b} concur with the results attained using various techniques \cite{hawking78,oshita2017,Ma2023try}.
\subsection{The \textit{4d}-flat Schwarzschild black hole in R\'enyi statistics}\label{renyi_section}

\paragraph{}We perform the same calculation for the four-dimensional asymptotically flat Schwarzschild black hole within the R\'enyi nonextensive formalism ($4d$ R\'enyi-Sch)\cite{Renyi:1959aa,Biro2013ytu,Czinner2016,Promsiri:2020jga,Promsiri:2021hhv}. One obtain \footnote[3]{For the R\'enyi-Sch black hole phase to exist the condition $\displaystyle\beta<\beta_{min}=\sqrt{\frac{4\pi}{\lambda}}$ must be satisfied \cite{Promsiri:2020jga}.},

\begin{equation}\label{eq_31}
S_{R}=\frac{1}{\lambda}\ln\left(1+\lambda\pi r_h^2\right), \quad M=\frac{r_h}{2} \quad \text{and}\quad  r_h=\displaystyle  \frac{2\sqrt{\pi}-\sqrt{4\pi- \beta^{2} \lambda}}{\sqrt{\pi} \beta \lambda} 
\end{equation}

Here, the nonextensivity is measured by the parameter $\alpha\equiv\lambda$ which is assumed to be small, $0<\lambda\ll1$, and accounts for the nonlocal and nonextensive nature of black holes. Through Eq.\eqref{eq_21}, one finds the following expression $(\beta\equiv \tau)$
\begin{align}\label{eq_32}
& L^\lambda_{cl}(\tau)\equiv L_{cl}(\tau)f^{'}_\lambda(\tau)= \displaystyle\sqrt{\frac{1}{\lambda \tau}-\frac{  \tau}{2\pi}}-\frac{\displaystyle1}{ \lambda  \tau }.
\end{align}
Here $ L_{cl}(\tau) $ is the Euclidean Lagrangian derived from the asymptotically flat Schwarzschild black hole metric, given by Eq.\eqref{eq_31b}, which is independent of the parameter $\lambda$. \textit{The nonextensivity is encoded in the generalized Wick's rotation represented by the function $f_\lambda(\tau)$}. A direct calculation of the asymptotically flat Schwarzschild black hole Euclidean action using the Hawking-Gibbon-York method confirms the expression of $ L_{cl}(\tau) $ as

\begin{equation}\label{eq_33}
L_{cl}(\tau)=-\frac{\beta}{8\pi}=-\frac{\tau}{8\pi}.
\end{equation}

Therefore, the derivative of the R\'enyi Wick's rotation function $f_\lambda(\tau)$ is found to be
\begin{equation} 
f^{'}_\lambda(\tau)=\frac{8 \pi}{\tau}\left(\frac{\displaystyle 1}{ \lambda  \tau }-\displaystyle\sqrt{\frac{1}{\lambda \tau}-\frac{  \tau}{2\pi}}\right).
\end{equation}

\paragraph{} For small $\lambda\ll1$, we get for $f_\lambda(\tau)$ and  $ \mathcal{I}^\lambda_{cl} $
\begin{equation}\label{eq_35}
f_\lambda(\tau)=\displaystyle \tau+\frac{\lambda \tau^{3}}{48 \pi}+O\left(\lambda^{2}\right),
\end{equation}

\begin{equation}\label{eq_36}
\mathcal{I}^\lambda_{cl}=\displaystyle \frac{\tau^{2}}{16 \pi}+\frac{\lambda \tau^{4}}{512 \pi^{2}} +O\left(\lambda^{2}\right).
\end{equation}
Consequently, the R\'enyi gravitational partition function, Eq.\eqref{eq_13}, is given by,

\begin{equation}\label{eq_38c}
Z_\lambda=\exp\left(\displaystyle -\frac{\beta^{2}}{16 \pi}-\frac{\lambda \beta^{4}}{512 \pi^{2}}\right)
\end{equation}
This R\'enyi partition function remains finite as long as $\lambda$ is positive, ensuring that no divergences occur. As $\lambda \rightarrow 0$, we recover the extensive Boltzmannian Wick's rotation function, $f_0(\tau) = \tau$. \textit{In summary, within this generalized Euclidean path integral formalism, to achieve the nonextensive R\'enyi statistical mechanics, one should perform the Wick rotation of the Lorentzian action such that:}
\begin{equation}\label{eq_40}
t\rightarrow-i\left(\displaystyle \tau+ \frac{\lambda \tau^{3}}{48 \pi}\right).
\end{equation}
\textit{This is in complete parallel with Wick's rotation to obtain the extensive Gibbs-Boltzmann statistical mechanics given by $ t\rightarrow -i\tau $.}

\paragraph{}Tsallis entropy\cite{Tsallis2023}, $S_T$, is connected to R\'enyi entropy $S_R$, through the relation,
\begin{equation}
   S_T \equiv\displaystyle\frac{\exp[{\displaystyle\left(1-q\right) S_R}]-1}{1-q},
\end{equation}
where the parameter $q=1-\lambda$. It is a simple calculation to compute the derivative of Wick's rotation function, $f^{'}_q(\tau)$, for the Tsallis statistics. One finds through Eq.\eqref{eq_21},
\begin{equation}
    f^{'}_q(\tau)=\displaystyle\frac{4\sqrt{\pi} \left[(1-q)^2 \tau ^4+8 \pi  (1-q) \tau ^2+32 \pi ^{3/2} \sqrt{4 \pi -(1-q) \tau ^2}-64 \pi ^2\right]}{(1-q)^2 \tau ^4 \sqrt{\displaystyle4\pi-(1-q) \tau ^2}},
\end{equation}
which for $q\rightarrow1$ becomes
\begin{align}
   &f^{'}_q(\tau)=\displaystyle 1-\frac{3  \tau ^4}{256 \pi ^2}(1-q)^2+O\left[(1-q) ^3\right],\label{eq_47e}\\
   \implies &  f_q(\tau)=\displaystyle \tau-\frac{3  \tau ^5}{1280 \pi ^2}(1-q)^2+O\left[(1-q) ^3\right].
\end{align}

\begin{figure}[!ht]
	\centering
	\includegraphics[scale=0.75]{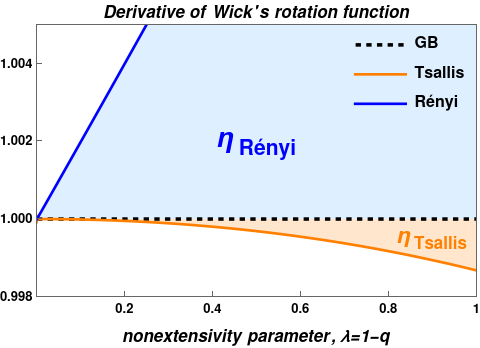}
 \vspace{-0.1cm}
	\caption{\footnotesize \it Comparing the derivatives of the Wick's rotation function of the Gibbs-Boltzmann, Tsallis, and R\'enyi statistics. The plot contrasts the nonextensivity nature of Tsallis and R\'enyi statistics for black hole thermodynamics. The shaded areas are a measure of the nonextensivity of the R\'enyi statistics $\displaystyle\eta_{\text{R\'enyi}}$ (Blue area) and the Tsallis statistics $\displaystyle\eta_{\,\text{Tsallis}}$ (Orange area). We fixed $\tau_0=1$.}
	\label{fig:fig1}
\end{figure}

Thus, the function, $f_q(\tau)$, differs only by a second order term in the nonextensivity parameter $\lambda=1-q$, from the Gibbs-Boltzmann Wick's rotation function $f_0(\tau)=\tau$. As depicted in Fig.\ref{fig:fig1}, comparing with the function $f_\lambda(\tau)$, Eq.\eqref{eq_40}, it is shown that R\'enyi (blue line) and Tsallis (orange line) statistics display different nonextensivity characters, with R\'enyi being more pronounced than Tsallis. Therefore, \textit{The present framework offers the possibility to contrast the nonextensive nature of different statistics based on their defining Wick's rotations}.

\subsection{The nonextensivity measure for statistics}
\paragraph{}In this section, we introduce a method to quantify the nonextensive nature of a given statistical model using the generalized Wick's rotation formalism. For this purpose, we define a measure, denoted as $\eta(S_\alpha)$, corresponding to a specific statistics $S_\alpha$, such that
\begin{equation}\label{eq_49e}
\eta(S_\alpha)=\underset{\tau\rightarrow\tau_0}{\lim}\,\bigg|\int_{0}^{1} \left(f^{'}_\alpha(\tau)-1\right) \,d\alpha\bigg|,
\end{equation}
where $f^{'}_\alpha(\tau)$ is the derivative with respect to $\tau$ of the Wick's rotation function of $S_\alpha$ and $\tau_0$ is a suitable nonzero imaginary time taken in support of $f^{'}_\alpha(\tau)$. Then\textit{ a statistics $S_{\alpha_{1}}$ is more nonextensive than another statistics $S_{\alpha_{2}}$} $\iff$ $\eta(S_{\alpha_{1}})>\eta(S_{\alpha_{2}})$. The definition Eq.\eqref{eq_49e} is stated such that $\eta(S_{GB})=0$. In Fig.\ref{fig:fig1}, we illustrated the geometrical meaning of the measure $\eta$ as the shaded areas. Using Eqs.\eqref{eq_40} and \eqref{eq_47e} one finds \footnote[4]{ A similar calculation for the nonextensive Kaniadakis statistics\cite{Kania2005,Kania2006} gives its nonextensivity measure as $\eta_{Kan}\simeq\displaystyle  \frac{1}{1536 \pi^{2}}$ },
\begin{equation}
\displaystyle \eta_{\text{\,R\'enyi}}\simeq\frac{1}{32\pi}
\quad \text{and}\quad \displaystyle \eta_{\text{\,Tsallis}}\simeq\frac{1}{256\pi^2},
\end{equation}
that is,  $\displaystyle\eta_{\,\text{\,R\'enyi}}>\eta_{\,\text{\,Tsallis}}$.The measure $\eta$ is independent of the specific nonextensivity parameter(s) on which a given statistical model may rely, providing a universal scale for assessing the degree of nonextensivity. Furthermore, this definition can be extended to accommodate statistics with multiple parameters. A straightforward extension involves replacing the single integral over one parameter with multiple integrals over all relevant parameters in Eq.\eqref{eq_49e}.
\paragraph{}A survey of the literature on the numerical quantification of nonextensivity reveals, to our knowledge, a lack of direct proposals in this area. However, Hanel and Thurner\cite{Hanel2011rrt,Hanel2011yyu} introduced a two-parameter asymptotic classification of generalized entropies by relaxing the fourth Shannon-Khinchin axiom\cite{Shannon1948,Khinchin1957}, building on earlier work by Tsallis\cite{Tsallis1988,Tsallis2013}. Moreover, numerous studies have established a connection between nonextensivity and complexity\cite{Grassberger1986,Beck2004,Tsallis2002opp,Bialek2001,Yamano2004}. In this context, the proposed measure of nonextensivity can also be interpreted as a measure of complexity.

\subsection{The generation of nonextensive statistics from Wick's rotation}
\paragraph{}Here, we investigate the reverse approach: given a specific Wick's rotation function and a Lorentzian action, it is possible to derive the corresponding statistical framework. We illustrate this procedure through a straightforward example and then proceed to a more generalized formulation.

\paragraph{}The second simplest generalized Wick's rotation one can postulate, after the Boltzmannian one, is of the form
\begin{equation}
t\rightarrow-i(1+\theta)\tau,
\end{equation}
where $\theta$ is a constant nonextensivity parameter. This rotation generates a nonextensive statistics, $S_\theta$, which can be computed through Eq.\eqref{eq_16}. Assuming a $4$-dimensional asymptotically flat Schwarzschild black hole metric, one writes a differential equation for $S_\theta$. Using Eq.\eqref{eq_16} and Eq.\eqref{eq_33}, we have
\begin{align}
M\frac{\partial S_\theta}{\partial M}-S_\theta=\mathcal{I}^\theta_{cl}=\displaystyle \frac{\beta^{2} \left(\theta + 1\right)}{16 \pi}.
\end{align}
However since $\displaystyle\beta\equiv\frac{\partial S_\theta}{\partial M}$, one finds that $S_\theta$ obeys a differential equation such as,
\begin{equation}\label{eq_53}
    \displaystyle \frac{ \left(\theta + 1\right)}{16 \pi}\left(\frac{\partial S_\theta}{\partial M}\right)^{2}-M\frac{\partial S_\theta}{\partial M}    +S_\theta=0.
\end{equation}
Eq.\eqref{eq_53} has the solution,
\begin{equation}
  S_\theta=\displaystyle \frac{4 \pi M^{2}}{\theta + 1}=\frac{\pi r_h^{2}}{\theta + 1},
\end{equation}
in which, we used the relation $r_h=2 M$ for the horizon radius of the asymptotically flat Schwarzschild black hole. It is clear that as the parameter $\theta \rightarrow0$, the entropy becomes that of GB, $S_\theta\rightarrow S_{GB}$. This $\theta$-statistics is the simplest generalization of the black hole GB-statistics one can contemplate. In the case of \textit{extreme nonextensivity}, $\theta\rightarrow\infty$, the black hole entropy $S_\theta$ vanishes and the black hole is in a \textit{perfectly-ordered} thermodynamic state with a unique micro-state.

\paragraph{}In general, for a given Wick's rotation defined by a function $f_\alpha(\tau)$ which produces a semi-classical Euclidean action $\mathcal{I}^\alpha_{cl}[\beta]$, the nonextensive statistics generated by such a Wick's rotation obeys the differential equation
\begin{equation}
   \bm{ M\frac{\partial S_\alpha}{\partial M}-S_\alpha-\mathcal{I}^\alpha_{cl}\left[\frac{\partial S_\alpha}{\partial M}\right]=0},
\end{equation}

where we used $\displaystyle\beta\equiv\frac{\partial S_\alpha}{\partial M}$.

\subsection{The \textit{4d}-flat Kerr black hole in R\'enyi statistics}

 We apply the formalism to the Kerr asymptotically flat black hole in R\'enyi statistics ($4d$ R\'enyi-Kerr). The expressions of the entropy, mass, and inverse temperature read as
\begin{equation}\label{eq_41}
S_{R}=\frac{1}{\lambda}\ln\left[1+\lambda\pi \left(r_h^2+a^2\right)\right], \quad M=\displaystyle \frac{r_{h}^{2}+a^{2}}{2 r_{h}}\quad \text{and}\quad \beta=\displaystyle \frac{4 \pi r_{h} \left( r_{h}^{2}+a^{2}\right)}{\left(r_{h}^{2}-a^{2} \right) \left[1+\pi \lambda \left( r_{h}^{2}+a^{2} \right) \right]}
\end{equation}

For \textit{small spin parameter} $a\ll r_h$, the horizon radius $r_h$ is expressed in terms of the inverse temperature $\beta$ such as
\begin{equation}
    r_h=\displaystyle \frac{\beta}{4 \pi}- \frac{8 \pi a^{2}}{\beta} +\frac{\lambda \beta^{3}}{64 \pi^{2}} + O\left(a^{4},\lambda^2\right).
\end{equation}
An analogous calculation to the preceding sections yields the expression for the Euclidean Lagrangian of R\'enyi-Kerr black hole as
\begin{align}\label{eq_42b}
&L^\lambda_{cl}(\tau,a)=\displaystyle - \frac{\tau}{8 \pi} + \lambda \left(- \frac{\tau^{3}}{128 \pi^{2}} + \frac{\tau a^{2}}{8}\right) + \frac{2 \pi a^{2}}{\tau}.
\end{align}

To retrieve Wick's rotation function, one needs the Euclidean Lagrangian for the asymptotically-flat Kerr black hole which can be found by putting $\lambda=0$ in Eq.\eqref{eq_42b}
\begin{equation}
    L^0_{cl}(\tau,a)=\displaystyle - \frac{\tau}{8 \pi}  + \frac{2 \pi a^{2}}{\tau}.
\end{equation}
Thus, applying Eq.\eqref{eq_21} and integrating with respect to $\tau$, the Wick's rotation function reads
\begin{align}
&f_\lambda(\tau)= \displaystyle \tau- \pi \lambda a^{2}  \tau+ \frac{\lambda \tau^{3}}{48 \pi}  + O\left(\lambda^2,a^{4}\right).\label{eq_44b}
\end{align}

We notice that Eq.\eqref{eq_44b} reduces to the function derived for the 4-dimensional R\'enyi-Schwarzschild black hole, as given in Eq.\eqref{eq_40}, in the limit $a \rightarrow 0$. Furthermore, Wick's rotation function exhibits dependence on the spin parameter $a$, which is intrinsic to the Lorentzian action of the Kerr metric. As demonstrated in Eq.\eqref{eq_41}, this dependence arises because, in the case of the Kerr black hole, the R\'enyi entropy itself becomes a function of $a$. 

\paragraph{}Lastly, the natural exponentiation of the Euclidean action,  gives the gravitational partition function for the $4d$ R\'enyi-Kerr,
\begin{equation}
    Z_\lambda[a]=\exp\left[\displaystyle -\frac{\tau^{2}}{16 \pi} + 2 \pi a^{2} \log{\left(\tau \right)} +\lambda\left( \frac{a^{2}\tau^{2} }{16}  +\frac{\tau^{4} }{512 \pi^{2}}\right)\right].
\end{equation}
In the limit of vanishing spin parameter, $a\rightarrow0$, we have the R\'enyi-Sch partition function, Eq.\eqref{eq_38c}.
\subsection{The \textit{4d}-flat Schwarzschild black hole in Barrow statistics}
In Barrow statistics\cite{Barrow2020}, the nonextensivity is quantified by the parameter $\Delta$. We have for the Barrow entropy, mass, and horizon radius of the $4d$ asymptotically flat Schwarzschild black hole ($4d$ Barrow-Sch)
\begin{equation}\label{eq_42}
S_{B}= \left(\displaystyle\pi r_{h}^{2}\right)^{\displaystyle1+\frac{\Delta}{2}}, \quad M=\displaystyle \frac{\displaystyle r_{h}}{2}\quad \text{and} \quad r_h=\displaystyle \left(\frac{\pi^{- \frac{\Delta-1}{2}} \beta}{2 \left(\Delta + 2\right)}\right)^{\displaystyle \frac{1}{\Delta + 1}}.
\end{equation}

Injecting these expressions in Eq.\eqref{eq_21}, we obtain the following expression for the Euclidean Lagrangian 
\begin{align}\label{eq_43}
&L^\Delta_{cl}(\tau)\equiv L_{cl}(\tau)f^{'}_\Delta(\tau)=\displaystyle-2^{-\frac{\Delta +2}{\Delta +1}} \left(\frac{\tau }{\pi ^{\frac{\Delta }{2}+1} (\Delta +2)}\right)^{\frac{1}{\Delta +1}}.
\end{align}

Just as before, $\displaystyle L_{cl}(\tau)$, is the Euclidean Lagrangian for the asymptotically flat Schwarzschild black hole metric, which is independent of the parameter $\Delta$ and given in Eq.\eqref{eq_33}. After integration with respect to imaginary time $\tau$, one gets the exact expression of Wick's rotation function for the Barrow statistics as
\begin{equation}
   f_\Delta(\tau)= \displaystyle2^{\frac{2 \Delta +1}{\Delta +1}}\pi(\Delta +1) \left(\frac{\tau}{\pi ^{\frac{\Delta }{2}+1} (\Delta +2)}\right)^{\frac{1}{\Delta +1}}.
\end{equation}
For small Barrow  parameter $\Delta\ll1$
\begin{equation}\label{eq_45}
     f_\Delta(\tau)=\tau +\frac{\tau}{2} \left[1 -  \ln \left(\frac{\tau^2 }{16\pi}\right)\right]\Delta+O\left(\Delta ^2\right).
\end{equation}

The generalized classical Euclidean action for Barrow statistics and the corresponding gravitational partition function read
\begin{align}\label{eq_46}
   &\mathcal{I}^\Delta_{cl}=\displaystyle \frac{\tau ^2 }{16 \pi}-\Delta \frac{\tau ^2 }{32 \pi} \ln \left(\frac{\tau^2}{16\pi}\right)+O\left(\Delta ^2\right),\\
   &Z_\Delta=\exp\left[\displaystyle -\frac{\tau ^2 }{16 \pi}+\Delta \frac{\tau ^2 }{32 \pi} \ln \left(\frac{\tau^2}{16\pi}\right)\right].
\end{align}
Once again, the extensive statistical mechanics are found in the limit of vanishing parameter $\Delta\rightarrow0$. \textit{As a summary, to obtain the nonextensive Barrow statistical mechanics, one should Wick rotate the Lorentzian action such as} 
\begin{equation}
    t\longrightarrow-i \left(\tau +\frac{\tau}{2} \left[1 -  \ln \left(\frac{\tau^2 }{16\pi}\right)\right]\Delta\right).
\end{equation}

In connection with the nonextensivity measure, applying Eq.\eqref{eq_49e} to the Barrow statistics reveals its expression to be
\begin{equation}
  \eta(S_\Delta) \simeq \displaystyle \frac{4\log{\left(2\sqrt[4]{\pi}\right)}-1}{4}.
\end{equation}

\subsection{The higher dimensional flat Schwarzschild black hole in R\'enyi statistics}

In this section, we extend section (\ref{renyi_section}) to higher dimensions and show how to compute Wick's rotation function in this case. We start with the generalization of Eqs.\eqref{eq_31} to $d$-dimensional spacetime

\begin{equation}\label{eq_65}
\hspace{-0.85cm}
S_{R}=\displaystyle \frac{1}{\lambda}\ln{\left(1+\displaystyle\frac{\lambda \pi^{\frac{d-1}{2}}  r_{h}^{d - 2}}{2 \Gamma\left(\frac{d-1}{2}\right)} \right)}, \; M=\displaystyle \frac{\pi^{\frac{d}{2} - \frac{3}{2}} r_{h}^{d - 3} \left(d - 2\right)}{8 \Gamma\left(\frac{d}{2} - \frac{1}{2}\right)} \; \text{ and}\; \beta=\displaystyle \frac{4 \pi^{\frac{3}{2}} r_{h} \Gamma\left(\frac{d}{2} - \frac{3}{2}\right)}{\pi^{\frac{d}{2}} \lambda r_{h}^{d-2} + 2 \sqrt{\pi}  \Gamma\left(\frac{d}{2} - \frac{1}{2}\right)}.
\end{equation}

Here, we express the inverse temperature $\beta$ in terms of the horizon radius $r_h$, as the direct inversion of this relationship is generally not feasible. The classical Euclidean action is then given by
\begin{equation}
   \mathcal{I}^{\lambda,d}_{cl} =\displaystyle\frac{\pi ^{d/2} (d-2) r_h^{d=2}}{(d-3) \left[\lambda \pi ^{d/2}   r_h^{d-2}+2 \sqrt{\pi} \Gamma \left(\frac{d-1}{2}\right) \right]}-\frac{1}{\lambda}\ln \left(\displaystyle1+\frac{\pi ^{\frac{d-1}{2}} \lambda  r_h^{d-2}}{2 \Gamma \left(\frac{d-1}{2}\right)}\right).
\end{equation}

The corresponding Euclidean Lagrangian, $L^{\lambda,d}_{cl}$, can be computed as
\begin{equation}
    L^{\lambda,d}_{cl}=-\frac{\partial \mathcal{I}^{\lambda,d}_{cl}}{\partial \tau}=-\frac{\partial \mathcal{I}^{\lambda,d}_{cl}}{\partial r_h}\;\frac{d r_h}{d\tau},
\end{equation}
then the Wick's rotation function $f^d_\lambda(\tau)$ is determined by
\begin{align}
    &\displaystyle L^{d}_{cl}\frac{df^d_\lambda}{d\tau}\equiv L^{\lambda,d}_{cl},\\
    \implies& \displaystyle \frac{df^d_\lambda}{d\tau}=\frac{L^{\lambda,d}_{cl}}{L^{d}_{cl}},\\
    \implies& \displaystyle\label{eq_60}\bm{\frac{df^d_\lambda} {dr_h}=\frac{L^{\lambda,d}_{cl}}{L^{d}_{cl}}\;\frac{d\beta}{dr_h}},\; 
 \text{since} \quad \beta\equiv\tau.
\end{align}
Here, $L^{d}_{cl}$, is the Euclidean Lagrangian of the $d$-dimensional asymptotically flat Schwarzschild black hole for $\lambda=0$. It generalizes Eq.\eqref{eq_33} as
\begin{equation}
    L^{d}_{cl}=-\displaystyle \frac{\pi^{\frac{d}{2} - \frac{3}{2}} r_{h}^{d - 3} \left(d - 2\right)}{8 \Gamma\left(\frac{d}{2} - \frac{1}{2}\right)}.
\end{equation}

From Eq.\eqref{eq_60}, we obtain
\begin{equation}
   \frac{df^d_\lambda} {dr_h}=\frac{2 \pi  \left( \Gamma \left(\frac{d-3}{2}\right) -\pi ^{\frac{d-1}{2}} \lambda  r_h^{d-2}\right)}{  \Gamma \left(\frac{d-1}{2}\right) +4 \pi ^{\frac{d-1}{2}} \lambda   r_h^{d-2}}.
\end{equation}
For small parameter $\lambda\ll 1$, a straightforward integration yields the R\'enyi Wick's rotation function in $d$-dimensions in terms of $r_h$ to be
\begin{equation}\label{eq_64}
    f^d_\lambda(r_h)=\frac{2}{(d-3)\Gamma \left(\frac{d-1}{2}\right)} \left[2 \pi  \Gamma \left(\frac{d-1}{2}\right) r_h-\pi ^{\frac{d+1}{2}} \lambda  r_h^{d-1}\right],
\end{equation}
and $r_h$ is expressed by Eq.\eqref{eq_65} as an implicit function of $\beta$. When $\lambda$ vanishes, one gets from Eqs.\eqref{eq_64} and \eqref{eq_65}
\begin{equation}
    f^d_0(r_h)=\frac{2 \pi \Gamma \left(\frac{d-3}{2}\right)}{\Gamma \left(\frac{d-1}{2}\right)}r_h =\frac{4\pi}{d-3}r_h\quad\text{and}   \quad\beta=\displaystyle \frac{2 \pi  \Gamma\left(\frac{d}{2} - \frac{3}{2}\right)}{   \Gamma\left(\frac{d}{2} - \frac{1}{2}\right)}r_{h}=\frac{4\pi}{d-3}r_h. 
\end{equation}
That is, $f^d_0(r_h)=\beta=\tau$. Thus, we arrive at the standard extensive Gibbs-Boltzmann Wick's rotation. This can be seen directly by inspection of Eq.\eqref{eq_60}; For $\lambda=0$, it reduces to
\begin{equation}
    \frac{df^d_0} {dr_h}=\frac{d\beta}{dr_h}\implies   \frac{df^d_0} {d\beta}=1\implies f^d_0(\beta)=\beta=\tau.
\end{equation}
After showing the applicability of the proposed formalism to some black hole/statistics combinations, we explore in the next section the R\'enyi/AdS equivalence.
\section{R\'enyi/AdS equivalence}\label{sec4}
\paragraph{}In the context of the conjectured equivalence between the \textit{4d AdS Schwarzschild black hole in extensive Gibbs-Boltzmann statistics (AdS-Sch)} and the \textit{4d nonextensive flat Schwarzschild black hole in R\'enyi statistics (R\'enyi-Sch)}, we propose to equate their partition functions. That is
	\begin{equation}\label{eq_76d}
	Z_{AdS_4}=Z_\lambda,
	\end{equation}
where $Z_{AdS_4}\equiv Z_0(\Lambda)$. Apart from strong connections revealed by recent studies\cite{Barzi2022,Barzi:2023mit,Barzi2024yan,Barzi2024yin}, which points to such equivalence, the present framework also provides support by remarking that their Euclidean actions, Eqs.\eqref{eq_29b} and \eqref{eq_36}, share the same functional form.  To first order in the parameters, $\Lambda$ and $\lambda$, they have the following partition functions

\begin{equation}
Z_{AdS_4}=\exp\left(-\frac{\beta^{2}}{16 \pi} + \frac{\Lambda \beta^{4}}{384 \pi^{3}}\right),
\end{equation}
and 
 \begin{equation}
Z_\lambda=\exp\left(\displaystyle -\frac{\beta^{2}}{16 \pi}-\frac{\lambda \beta^{4}}{512 \pi^{2}}\right).
\end{equation}
Thus, applying \eqref{eq_76d} gives the relation
\begin{equation}\label{eq_72b}
    \Lambda=-\frac{3\pi}{4}\lambda+O\left(\lambda^2\right),
\end{equation}
which aligns with previous findings obtained using the \textit{Hamiltonian approach to thermodynamics}\cite{Barzi:2023mit}. Thermodynamically, this indicates that a $4d$ AdS-Schwarzschild black hole is equivalent to a $4d$ R'enyi-Schwarzschild black hole, provided that Eq.\eqref{eq_72b} holds.

\paragraph{}In the context of the $AdS_5/CFT_4$ duality, The GKP-Witten relation\cite{GKP98,Witten98} proposes the equality of the partition functions of string theory
on $5$-dimensional $AdS_5$ spacetime and the $4$-dimensional conformally invariant gauge theory such as,
\begin{equation}\label{eq_73}
Z_{AdS_5}=Z_{CFT_4}.
\end{equation}
In the large-$N_c$ limit for the gauge side of Eq.\eqref{eq_73}, one can use the saddle-point-approximation for the gravity side, which gives $Z_{AdS_5}$ in terms of the Euclidean classical action. Consequently, a parameter dictionary can be derived for this gauge/gravity duality such as
\begin{equation}\label{eq_74}
    N_c^2=\frac{\pi}{2 G_5}\left(-\frac{6}{\Lambda_5 }\right)^{\frac{3}{2}}
\end{equation}
where $N_c$ is the number of charges of the gauge group and $G_5$ is Newton's constant in $5$-dimensional spacetime. The gravitational partition functions of the $5d$ AdS-Sch and R\'enyi-Sch black holes are calculated as
\begin{equation}
 Z_{AdS_5}=\displaystyle \exp\left( -\frac{\beta^{3}}{32 \pi} + \frac{\Lambda_5 \beta^{5}}{256 \pi^{3}}\right)\quad\text{and}\quad   Z^5_\lambda= \exp\left( - \frac{\beta^{3}}{32 \pi}-\frac{\lambda_5\beta^{6}}{512 \pi^{2}}\right) 
\end{equation}

Thus, an equivalence holds between these two thermodynamic systems if,
\begin{equation}\label{eq_76}
     \Lambda_5=-\frac{\pi}{2}\lambda_5\beta+O\left(\lambda_5^2\right)
\end{equation}
An interesting interpretation of Eq.\eqref{eq_76} comes from considering the extended phase space. It is usual to treat the cosmological constant $\Lambda$ as the thermodynamic pressure $P$\cite{Dolan2011,Belhaj2012},
\begin{equation}\label{eq_77}
    P=-\frac{\Lambda_5}{8\pi}.
\end{equation}
Similarly, one defines the R\'enyi pressure\cite{Barzi2022}, $P_R$ in $5$-dimensional spacetime as,

\begin{equation}\label{eq_78}
    P_R=\displaystyle \frac{3 \pi }{16}\lambda_5 r_{h}.
\end{equation}
From the expression of $\beta$ in terms of the horizon radius $r_h$, Eq.\eqref{eq_41}, it is straightforward to obtain an inverted expression in five dimensions and for small $\lambda$ such as,
\begin{equation}
    r_h=\displaystyle  \frac{\beta}{2 \pi}+O\left(\lambda_5\right)
\end{equation}
By substitution in \eqref{eq_78}, we have for the R\'enyi pressure,
\begin{equation}\label{eq_80}
    P_R=\displaystyle \frac{3 }{32}\lambda_5 \beta+O\left(\lambda_5^2\right).
\end{equation}
A quick comparison of Eqs.\eqref{eq_76} and \eqref{eq_80}, taking into consideration Eq.\eqref{eq_77} gives,
\begin{align}
    &\Lambda_5=-8\pi P=-\frac{16\pi}{3}P_R,\\
   \implies& \bm{P=\frac{2}{3}P_R}.
\end{align}
In this respect, Eq.\eqref{eq_74} can be reformulated in terms of pressure as
\begin{equation}
    N_c^2=\frac{\pi}{2G_5}\left(\frac{3}{4\pi P }\right)^{\frac{3}{2}}=\frac{\pi}{2G_5}\left(\frac{9}{8\pi P_R}\right)^{\frac{3}{2}}.
\end{equation}
Also, in terms of length scales, one has the usual AdS spacetime radius $\displaystyle L=\sqrt{\frac{3}{4\pi P}}$ and the R\'enyi length scale $\displaystyle L_\lambda=\sqrt{\frac{9}{8\pi P_R}}$, then
\begin{equation}
    N_c^2=\frac{\pi}{2 G_5}L^3=\frac{\pi}{2 G_5}L_\lambda^3.
\end{equation}
In view of the {\it R\'enyi}$\,_5$/$AdS_5$ equivalence\footnote[8]{This equivalency can be easily extended to all $d$-dimensional spacetimes {\it R\'enyi}$_d$/$AdS_d$, $d>3$. In particular, for $d=4$ spacetime, $\displaystyle P_R=\frac{3\lambda}{32.}$, consequently one has $P=P_R$.}, we see that the expression of the $AdS_5/CFT_4$ correspondence induces the duality {\it R\'enyi}$_5$/$CFT_4$  with the dictionary,
\begin{equation}
    N_c^2=\frac{\pi}{2 G_5}L_\lambda^3 \quad \text{and} \quad \lambda_{t}=\left(\frac{L_\lambda}{l_s}\right)^4.
\end{equation}

Where $\mathcal{\lambda}_t$ is the 't Hooft parameter and $l_s$ is the string length. \textit{The two sides of the duality in this new form are both flat theories}. A further investigation of this reformulation is warranted.
 \section{Conclusion}\label{sec5}
 In this study, we generalize the Euclidean path integral to derive nonextensive thermodynamics from Lorentzian action through a generalized Wick's rotation. Within this framework, nonextensive and extensive statistics are placed on an equal basis as Wick's rotation of the same action. This in essence decouples the Lorentzian action which determines the equations of motion of a given system from the choice of the statistics to apply to it. Thus recovering the freedom to choose the statistics independent from the action. Besides, one can formally implement any statistics as a rotation of the Lorentzian action provided a Wick's rotation function exists instead of imposing the statistics in an ad hoc manner at a later stage. Additionally, by choosing different Wick's rotation functions and analyzing their deviations from extensivity, new statistical models can be generated. Specifically, constructing a Wick's rotation function that deviates from the extensive Boltzmannian form, $f_0(\tau) = \tau$, either slightly or significantly, allows for the production of weakly or strongly nonextensive statistics. The most significant conceptual advantage of this proposed generalization is its ability to incorporate nonextensivity within the linear framework of quantum mechanics, which has traditionally been closely tied to the extensive Gibbs-Boltzmann statistics through the standard Wick's rotation.

\paragraph{} We proceeded to apply the generalized Euclidean path integral in the saddle point approximation to black hole thermodynamics and computed Wick's rotation functions for a variety of black holes under different statistics. We showed that R\'enyi statistics is strongly nonextensive in contrast to the Tsallis one by introducing a novel measure $\eta$ to quantify the nonextensive character of any statistics. Thus providing an order in the set of all nonextensive statistical mechanics. Furthermore, we examined the \textit{R\'enyi/AdS} equivalence which we translated to a \textit{R\'enyi/CFT} correspondence whenever the \textit{AdS/CFT} one holds. Moreover, in this formulation, we have flat theories on both sides of the duality. Also, this suggests a novel type of correspondence, namely, \textit{a gauge/statistics duality}. The investigation of such a theme is one of the main motivations of this study and we intend to deepen our inquiry to reach further insights.
\bibliographystyle{unsrt}
\bibliography{Euclidean.bib} 
\end{document}